\documentclass{article}
\usepackage{amsfonts,amssymb}
\usepackage[noblocks]{authblk}
\usepackage{amsthm}
\usepackage{geometry,algorithm}  
\usepackage{algorithmic}
\usepackage{amsmath} 
\usepackage{cite}  
\usepackage{graphicx}  
\usepackage{caption}
\usepackage{subfigure}
\usepackage{subfigure}
\usepackage{subfigure}
\usepackage{float}
\usepackage{booktabs}
\usepackage{multirow}
\usepackage{threeparttable}
\usepackage{caption}
\usepackage[utf8]{inputenc}
\usepackage[T1]{fontenc}
\usepackage{booktabs}
\usepackage{array, caption, threeparttable}
\usepackage[font=small,labelfont=bf,labelsep=none]{caption}
\newtheorem{theorem}{Theorem}
\usepackage{tikz}
\newcommand*{\circled}[1]{\lower.7ex\hbox{\tikz\draw (0pt, 0pt)%
  circle (.5em) node {\makebox[1em][c]{\small #1}};}}
\begin{document}
\renewcommand{\thefootnote}{\fnsymbol{footnote}}
\captionsetup[figure]{labelfont={bf},name={Fig },labelsep=period}
\title{\textbf {Effects of measurement dependence on  1-parameter family of Bell tests} }
\author[1,2]{Fen-Zhuo Guo}
\author[1,2]{Ze-Tian Lv}
\author[1,2]{Shi-Hui Wei}
\author[1]{Qiao-Yan Wen}
\affil[1]{
	\emph{State Key Laboratory of Networking and Switching Technology, Beijing University of  Posts and Telecommunications, Beijing 100876, China}}
\affil[2]{
	 \emph{School of Science, Beijing University of Posts and Telecommunications, Beijing 100876, China}}
\maketitle

{\noindent\textbf {Abstract}\quad}{Most quantum information tasks based on Bell tests relie on the assumption of measurement independence. However, it is difficult to ensure that the assumption of measurement independence is always met in experimental operations, so it is crucial to explore the effects of relaxing this assumption on Bell tests. In this paper, we discuss the effects of relaxing the assumption of measurement independence on 1-parameter family of Bell (1-PFB) tests. For both general and factorizable input distributions, we establish the relationship among measurement dependence, guessing probability, and the maximum value of 1-PFB correlation function that Eve can fake. The deterministic strategy when Eve fakes the maximum value is also given. We compare the unknown information rate of Chain inequality and 1-PFB inequality, and find the range of the parameter in which it is more difficult for Eve to fake the maximum quantum violation in 1-PFB inequality than in Chain inequality.
\section{Introduction}
\label{intro}
As one of the most striking properties of quantum mechanics, quantum nonlocality\cite{quantum nonlocality1,quantum nonlocality2,quantum nonlocality3,quantum nonlocality4} plays a prominent role in device independent (DI) quantum
information theory such as quantum key distribution\cite{QKD1,QKD2},  random generation\cite{robust,QRNG1,QRNG2} and entanglement certification\cite{entanglement certification1,entanglement certification2}. In the DI framework, we don't need to consider specific internal structures but only focus on the correlation between input and output.

Bell tests\cite{Bell test1,Bell test2,Bell test3,Bell test4}, as an original tool for observing quantum nonlocality, can detect the correlation between input and output. Generally, the participants involved in the Bell tests first randomly select the input, and then produce the output through the measurement performed on the physical system. 
In the DI scenario, the violation of Bell inequality is commonly used in analysis of quantum theory protocols. For example, for a general random number generator, an eavesdropper (Eve) may choose a predetermined string as the output in which instance we cannot ensure whether the output contains true randomness. DI quantum random number generator based on Bell tests can avoid the above vulnerability, in which Eve can not fake the violation of Bell inequality, because her predetermined output is independent of the input randomly selected by the participants.

The initial derivation of Bell inequality is based on the assumption that participants randomly select the input (i.e., measurement independence)\cite{QRNG5}. 
In actual experiments, it is difficult to ensure that the assumption is satisfied. For example, Eve may control the input of the device, such that the participants cannot randomly select the input, in other words, Eve obtains part information of the input. In this case, Eve can choose the appropriate string as the output according to the obtained input information, thus forging the violation of the Bell inequality. In view of this, the study of relaxing the assumption of measurement independence\cite{QRNG4,QRNG5,QRNG6,QRNG7,QRNG8,QRNG9,QRNG10,QRNG11,QRNG12,QRNG13,QRNG14,QRNG15,QRNG16} has attracted widespread attention. For convenience, we will relax this assumption and uniformly call it measurement dependence. Koh et al. \cite{QRNG6} studied the effects of  measurement dependence on the Clauser-Horne-Shimony-Holt (CHSH) Bell inequality\cite{CHSH}. They gave the relationship among measurement dependence, guessing probability, and the maximum violation value of CHSH Bell inequality that Eve can fake. Pütz et al. \cite{QRNG9} proved that an arbitrarily small amount of measurement independence is sufficient to demonstrate quantum nonlocality. Refs. \cite{QRNG8} and \cite{QRNG10} both gave the Eve's optimal strategy for forgery, and then gave the maximum value of the CHSH Bell correlation function that Eve can fake under the general input distribution and the factorizable input distribution respectively. Li et al. \cite{QRNG12} explored the effects of measurement dependence on the
generalized CHSH Bell tests in both single-run and multiple-run scenarios, they found that it is more difficult for Eve to fake a violation in the generalized CHSH Bell tests in some special
cases by comparing with the simplest CHSH Bell
tests. Huang et al. \cite{tilted} investigated the effects of measurement dependence on the tilted CHSH Bell inequality under different input distributions.

All the above attempts to characterize quantum nonlocality are based on Bell inequality. 
In the case where each party has two possible 2-outcome measurements (i.e., the case 2222), there is only an equivalence named CHSH Bell inequality (i.e., a special Chain inequality with two measurement settings)\cite{Chain,xin Bell,9}. In the case where each party has three possible 2-outcome measurements (i.e., the case 3322), there is another new class of inequality $ I_{3322} $ Bell inequality besides Chain inequality, and the $ I_{3322} $ Bell inequality can detect the entanglement of more states \cite{xin Bell}.  Kaniewski \cite{family} presented $ I_{3322} $ with parameter $ \alpha $ called 1-PFB inequality which are maximally violated by multiple inequivalent quantum realizations, and showed that it can be used to robustly self-test quantum state and certificate randomness. So, some key questions arise: What will happen on 1-PFB tests if we relax the assumption of measurement independence? Can we extract true randomness in the process of relaxing the assumption of measurement independence? Also is it more difficult for Eve to fake the violation in 1-PFB inequality than in Chain inequality with three measurement settings?

Inspired by the works in Refs. \cite{QRNG12,tilted}, we study the effects of measurement dependence on 1-PFB inequality under general and factorizable input distributions, respectively. In both cases, we use the flexible upper and lower bound of measurement dependence to establish the relationship among measurement dependence, guessing probability, and the maximum value of 1-PFB inequality that Eve can fake. We also give the strategy for Eve to forge the maximum violation value. At the same time, we find that there exists true randomness in certain conditions. We briefly give the similar relationship on Chain inequality where each party has three measurement settings. By comparing the conclusions of 1-PFB inequality and Chain inequality, we find that in some circumstances, it is more harder for Eve to fake maximum quantum violation in 1-PFB inequality than in Chain inequality, but in other cases, the results are reversed.

The structure of this paper is as follows: We will briefly introduce the relevant knowledge of Bell inequalities and measurement dependence in Sec. \ref{sec2}. In Subsec. \ref{sec31} and Subsec. \ref{sec32}, the relationship among measurement dependence, guessing probability, and the maximum value of 1-PFB correlation function that Eve can fake under different input distributions will be given, and a similar relationship will also be given in Chain inequality with three measurement settings. We will compare 1-PFB inequality with Chain inequality in Subsec. \ref{sec33}. Finally, we will conclude our results in Sec. \ref{sec4}.
\section{Preliminaries}
\label{sec2}
In this section, we introduce the knowledge of Bell inequalities and measurement dependence \cite{QRNG12,tilted}.
\subsection{Bell inequalities with three measurement settings}
\label{sec:2}
In the simplest scenario,  two-party, Alice and Bob, have three measurement settings with two outputs respectively. 
The measurement settings of Alice and Bob are marked by $ X_j $ and $ Y_k $ respectively, where $ j $, $ k\in\{0,1,2\} $. The outputs of Alice and Bob are marked by $ a $ and $ b $ respectively, where $ a $, $ b\in\{0,1\} $. After running the Bell experiment many times, conditional probability  distribution $ p(a,b|X_j,Y_k) $ will be obtained. There are two linear constraints based on probability distribution for the Bell tests with three measurement settings. The two linear constraints include the Chain inequality with three measurement bases and the $ I_{3322} $ Bell inequality. 

The Chain inequality where each party
has three measurement settings\cite{CHSH,Chain} is defined by
\begin{equation}
	\label{I1}
	I_{Chain}=\left \langle X_0Y_0 \right \rangle+\left \langle X_1Y_1 \right \rangle+\left \langle X_2Y_2 \right \rangle+\left \langle X_1Y_0 \right \rangle+\left \langle X_2Y_1 \right \rangle-\left \langle X_0Y_2 \right \rangle \le I_C,
\end{equation}
where $ \left \langle X_jY_k \right \rangle = \displaystyle\sum_{a,b}p(a=b|X_j,Y_k)-p(a\ne b|X_j,Y_k) $. For classical theory, it is easy to check that for equation (\ref{I1}) the maximum value is $ I_C=4 $. In the quantum mechanic, $ p(a,b|X_j,Y_k)=\mathrm{tr}(\rho M^a_j\bigotimes M^b_k) $, where state $ \rho $ is shared between Alice and Bob, and $ M^a_j$ $(M^b_k) $ represents the measurement operators of Alice (Bob). In this case, equation (\ref{I1}) can reach the maximum value 3$\sqrt{3}$. For no-signaling theory, the maximum value reaches $ 6 $ in equation (\ref{I1}). 

As for $ I_{3322} $ Bell inequality, its more general form called 1-PFB inequality has been proposed in Ref. \cite{family} 
\begin{equation}
	\label{B1}
	I^\alpha_{3322}=\left \langle X_0Y_0 \right \rangle+\left \langle X_0Y_1 \right \rangle+\alpha\left \langle X_0Y_2 \right \rangle+\left \langle X_1Y_0 \right \rangle+\left \langle X_1Y_1 \right \rangle-\alpha\left \langle X_1Y_2 \right \rangle+\alpha\left \langle X_2Y_0 \right \rangle-\alpha\left \langle X_2Y_1 \right \rangle \le I^\alpha_{3322,C},
\end{equation}
where $ \alpha\in[0,2] $ (the relevant definitions are the same as above). Particularly, when $ \alpha=1 $, equation (\ref{B1}) is $ I_{3322} $ Bell inequality \cite{xin Bell,family2}. Obviously, $ I^\alpha_{3322,C}=\max\{4,4\alpha\} $ is the classical bound of $ I^\alpha_{3322} $. For quantum theory, the upper bound of $ I^\alpha_{3322} $ is $ I^\alpha_{3322,Q}=4+\alpha^2 $.  Similarly, for no-signaling theory, denote the corresponding upper bound as $ I^\alpha_{3322,NS}=4+4\alpha $. $ I^\alpha_{3322} $ Bell inequality satisfying $ I^\alpha_{3322,C}=I^\alpha_{3322,Q} $ cannot be used to certify quantum properties, so we only consider the case with $ I^\alpha_{3322,C} < I^\alpha_{3322,Q} $ (i.e., $ \alpha\in(0,2) $). 

In this paper, we will focus on discussing the effects of measurement dependence on 1-PFB inequality under different input distributions, and use Chain inequality as a comparison.
\subsection{Measurement dependence}
\label{sec22}
Generally, there are two black boxes, Alice and Bob. The inputs of Alice and Bob are marked by $ X_j $ and $ Y_k $ respectively (with $ j,k \in\{0,1,2\} $), and the outputs are
marked by $ a $ and $ b $ respectively (with $ a,b \in\{0,1\} $).
After many runs, the probability distribution of output conditioned on input will be obtained
\begin{equation}
	\label{E2}
	p(a,b|X_j,Y_k)=\dfrac{\sum_{\lambda}p(\lambda)p(X_j,Y_k|\lambda)p(a,b|X_j,Y_k,\lambda)}{p(X_j,Y_k)},
\end{equation}
where $ \lambda $ is a local hidden variable or strategy. According to local correlations theory\cite{local}, $ p(a,b|X_j,Y_k,\lambda) $ can be decomposed into $ p(a|X_j,\lambda)p(b|Y_k,\lambda) $, so equation (\ref{E2}) is further written as
\begin{equation}
	\label{E3}
	p(a,b|X_j,Y_k)=\dfrac{\sum_{\lambda}p(\lambda)p(X_j,Y_k|\lambda)p(a|X_j,\lambda)p(b|Y_k,\lambda)}{p(X_j,Y_k)}.
\end{equation}
To make our results more powerful,
we adopt the method in Ref. \cite{tilted} to define the parameters $ P $ and $ S $ as flexible upper and lower bound of probabilities respectively for a set of selected specific measurement settings 
\begin{equation}
	\begin{split}	
		\label{E4}
		\max p(X_j,Y_k|\lambda) = P,\\
		\min p(X_j,Y_k|\lambda) = S,
	\end{split}
\end{equation}
for any $ \lambda $.

Since each party has three measurement settings, we can easily get $ P\in[\dfrac{1}{9},1] $ and $ S\in[0,\dfrac{1}{9}] $. Then, we analyze the situation when $ P $ and $ S $ take different values.\\
(1) $ P =\dfrac{1}{9} $ or $ S =\dfrac{1}{9} $, it means that Eve will not obtain information with the local hidden variables $ \lambda $, that is, the inputs are entirely random. \\
(2) $ P\in(\dfrac{1}{9},1) $ or $ S\in(0,\dfrac{1}{9}) $, it means that Eve can obtain part of the input information through the local hidden variables $\lambda$.\\
(3) $ P = 1 $, it means that Eve can obtain all the input information by using the local hidden variables $\lambda$, that is, Eve completely controls the input by using a deterministic strategy.

Here, we describe the predictability of outputs as guessing
probability $ G $ with \cite{tilted,QRNG6,QRNG12}
\begin{equation}
	\label{E5}
	G = \sum_{\lambda}p(\lambda)G(\lambda),
\end{equation}
where $ G(\lambda) = \displaystyle\max_{a,b,j,k}\{p(a|X_j,\lambda),p(b|Y_k,\lambda)\} $. For a given hidden variable $\lambda$, $ G(\lambda) $ represents the upper bound of the probability that Eve guesses the output. $ G = \dfrac{1}{2}$ $(G = 1) $ implies that Eve has an entirely indeterministic (deterministic) strategy. 
\section{\textbf{The effects under different input distributions}}
\label{sec3}
In this paper, we discuss the general input distribution and factorizable input distribution \cite{tilted,QRNG12}. Specifically, the factorizable input distribution is formulated as
\begin{equation}
	\label{E61}
	p(X_j,Y_k|\lambda)= p(X_j|\lambda) p(Y_k|\lambda).
\end{equation}
The distribution $ p(X_j,Y_k|\lambda) $ that cannot be expressed in equation (\ref{E61}) is called general input distribution. Subsequently, we will discuss the effects of measurement dependence under different input distributions and compare the effects of measurement dependence on different inequalities. 
\subsection{The general input distribution}
\label{sec31}
Firstly, we give the relationship among measurement dependence, guessing probability, and 
the maximum value of 1-PFB correlation function that Eve can fake under the general input distribution.
\begin{theorem}
	\label{th1}
	The maximum value of 1-PFB correlation function that Eve can fake, $ I^\alpha_{3322}(G,S,P) $, for any no-signaling model with $ p(X_j,Y_k) = \dfrac{1}{9} $, is
	
	\begin{equation}
		\label{E7}
		\\I^\alpha_{3322}(G,S,P)=
		\begin{cases}
			4\alpha + 4 - 36S,&\text{$8P + S \geq 1$},\\
			4\alpha + 4 - 36(2G - 1)(1 - 8P),&
			\text{$8P + S < 1$},
		\end{cases}\
	\end{equation}
	where the upper bound and lower bound of measurement dependence are $ P $ and $ S $, and guessing probability is characterized by $ G $.
\end{theorem}
\indent \textbf{{Proof}}
Based on flexible upper and lower bounds of the measurement dependence given in equation (\ref{E4}), we get
\begin{equation}
	\label{E8}
	P\le p^{'}(X_j,Y_k|\lambda)\le S,\\\ \forall j,k,\lambda,
\end{equation}
where $ j, k\in\{0,1,2\} $. For any set of elements $ (X_j, Y_k, \lambda) $, we let
\begin{equation}
	\label{E9}
	p(X_j,Y_k|\lambda)=\dfrac{p^{'}(X_j,Y_k|\lambda)-S}{1-9S}.
\end{equation}
According to equations (\ref{E8}) and (\ref{E9}), we get
\begin{equation}
	\label{E10}
	0\le p(X_j,Y_k|\lambda)\le \frac{P-S}{1-9S},\\\ \forall j,k,\lambda.
\end{equation}
We can prove
\begin{equation}
	\begin{aligned}
		\label{E100}
		\displaystyle\sum_{j,k}p(X_j,Y_k|\lambda)=& \sum_{j,k}\dfrac{p^{'}(X_j,Y_k|\lambda)-S}{1-9S}\\
		=& \dfrac{\sum_{j,k}[p^{'}(X_j,Y_k|\lambda)-S]}{1-9S}\\
		=& \dfrac{\sum_{j,k}p^{'}(X_j,Y_k|\lambda)-9S}{1-9S}\\
		=& 1.
	\end{aligned}
\end{equation}

Next, we estimate $ \left\langle X_jY_k \right\rangle $ based on the probability distribution. Let $ p_{A}(-1|X_j,\lambda)=m_j $, $ p_{B}(-1|Y_k,\lambda)=n_k $, $ p(-1,-1|X_j,Y_k,\lambda)=c_{j,k} $, we get
\begin{equation}
	\begin{split}
		\label{E11}
		p(-1,1|X_j,Y_k,\lambda)=m_j-c_{j,k},\\
		p(1,-1|X_j,Y_k,\lambda)=n_k-c_{j,k},\\
		p(1,1|X_j,Y_k,\lambda)=1+c_{j,k}-m_j-n_k,
	\end{split}
\end{equation} 
so
\begin{equation}
	\label{E12}
	p(a,b|X_j,Y_k,\lambda)\in \{c_{j,k},m_j-c_{j,k},1+c_{j,k}-m_j-n_k\}.
\end{equation}
As we know, $ c_{j,k} $ satisfies
\begin{equation}
	\label{15eq}
	\max\{0,m_j+n_k-1\}\le c_{j,k}\le \min\{m_j,n_k\},
\end{equation}
where
\begin{equation}
	\label{16eq}
	\begin{aligned}
		\min\{x,y\}=\frac{1}{2}[x+y-|x-y|],\\ \max\{x,y\}=\frac{1}{2}[x+y+|x-y|].
	\end{aligned}
\end{equation}
According to the previous definition of $ \left\langle X_jY_k \right\rangle $, we gain
\begin{equation}
	\label{17eq}
	\begin{aligned}
		\left \langle X_jY_k \right \rangle =& \displaystyle\sum_{a,b}p(a=b|X_j,Y_k)-p(a\ne b|X_j,Y_k)\\=&1+4c_{j,k}-2(m_j+n_k).
	\end{aligned}
\end{equation}
By applying equations (\ref{15eq}), (\ref{16eq}) and (\ref{17eq}), the bound of $ \left\langle X_jY_k \right\rangle $ is given by
\begin{equation}
	\label{E13}
	2|m_j+n_k-1|-1\le\left\langle X_jY_k \right\rangle \le 1-2|m_j+n_k|.
\end{equation}
Hence, $ \overline{I}^\alpha_{3322} $ can be written as
\begin{equation}
	\begin{aligned}
		\label{E14}
		\overline{I}^\alpha_{3322}
		=&\displaystyle\sum_{\lambda}[p(\lambda|X_0Y_0)\left \langle X_0Y_0 \right\rangle+p(\lambda|X_0Y_1)\left \langle X_0Y_1 \right\rangle+\alpha p(\lambda|X_0Y_2)\left \langle X_0Y_2 \right\rangle+p(\lambda|X_1Y_0)\left \langle X_1Y_0 \right\rangle\\&+p(\lambda|X_1Y_1)\left \langle X_1Y_1 \right\rangle-\alpha p(\lambda|X_1Y_2)\left \langle X_1Y_2 \right\rangle+\alpha p(\lambda|X_2Y_0)\left \langle X_2Y_0 \right\rangle-\alpha p(\lambda|X_2Y_1)\left \langle X_2Y_1 \right\rangle]\\\le&\displaystyle\sum_{\lambda}[p(\lambda|X_0Y_0)(1-2|m_{0}-n_{0}|)+p(\lambda|X_0Y_1)(1-2|m_{0}-n_{1}|)+\alpha p(\lambda|X_0Y_2)(1-2|m_{0}-n_{2}|)\\&+p(\lambda|X_1Y_0)(1-2|m_{1}-n_{0}|)+p(\lambda|X_1Y_1)(1-2|m_{1}-n_{1}|)-\alpha p(\lambda|X_1Y_2)(2|m_{1}+n_{2}-1|-1)\\&+\alpha p(\lambda|X_2Y_0)(1-2|m_{2}-n_{0}|)-\alpha p(\lambda|X_2Y_1)(2|m_{2}+n_{1}-1|-1)]\\\le&4+4\alpha-2\displaystyle\sum_{\lambda}[p(\lambda|X_0Y_0)|m_{0}-n_{0}|+p(\lambda|X_0Y_1)|m_{0}-n_{1}|+\alpha p(\lambda|X_0Y_2)|m_{0}-n_{2}|+p(\lambda|X_1Y_0)\\&|m_{1}-n_{0}|+p(\lambda|X_1Y_1)2|m_{1}-n_{1}|+\alpha p(\lambda|X_1Y_2)|m_{1}+n_{2}-1|+\alpha p(\lambda|X_2Y_0)|m_{2}-n_{0}|+\alpha p(\lambda|X_2Y_1)\\&|m_{2}+n_{1}-1|]-2(\alpha-1)\displaystyle\sum_{\lambda}[p(\lambda|X_0Y_2)|m_{0}-n_{2}|+p(\lambda|X_1Y_2)|m_{1}+n_{2}-1|+p(\lambda|X_2Y_0)|m_{2}-n_{0}|\\&+p(\lambda|X_2Y_1)|m_{2}+n_{1}-1|]\\\le&4+4\alpha-36(2G-1)\displaystyle\sum_{\lambda}p(\lambda)\min p(X_j, Y_k|\lambda),
	\end{aligned}
\end{equation}
when $ m_{2}+n_{1}=1 $, equality holds.

Based on the results of Ref. \cite{QRNG6}, we analyze the value of $ \min p(X_j, Y_k|\lambda) $:\\
(1) If $ P\ge\frac{1}{8} $, we find that $ \min p(X_j, Y_k|\lambda)=0 $.\\
(2) If $ \frac{1}{9}\le P\le\frac{1}{8} $, let $ p(X_j, Y_k|\lambda)=P $, where $ (j,k)\ne(j_1,k_1) $, so $ \min p(X_j, Y_k|\lambda)=p(X_{j_1}, Y_{k_1}|\lambda)=1-8P $. Then equation (\ref{E14}) can be simplified to
\begin{equation}
	\label{E15}
	\\\overline{I}^{\alpha}_{3322}(G,P)=
	\begin{cases}
		4\alpha + 4, &\text{$ P \geq \dfrac{1}{8}$},\\
		4\alpha + 4 - 36(2G - 1)(1 - 8P),&
		\text{$ \dfrac{1}{9}\le P < \dfrac{1}{8} $}.
	\end{cases}\
\end{equation}
On this basis, $ I^\alpha_{3322}(G,S,P) $ can be described as
\begin{equation}
	\begin{aligned}
		\label{E16}
		I^\alpha_{3322}
		\leq&4 + 4\alpha-2\sum_\lambda[p(\lambda|X_0Y_0)(|m_0-n_0|)+p(\lambda|X_0Y_1)(|m_0-n_1|)+\alpha{p(\lambda|X_0Y_2)(|m_0-n_2|)}\\
		&+p(\lambda|X_1Y_0)(|m_1-n_0|)+p(\lambda|X_1Y_1)(|m_1-n_1|)+p(\lambda|X_1Y_2)(|m_1+n_2-1|)+\alpha\\
		&{p(\lambda|X_2Y_0)(|m_2-n_0|)}+p(\lambda|X_2Y_1)(|m_2+n_1-1|)]-2(\alpha-1)\sum_\lambda[{p(\lambda|X_0Y_2)}(m_0-n_2)\\
		&+p(\lambda|X_1Y_2)(|m_1+n_2-1|)+\alpha p(\lambda|X_2Y_0)(|m_2-n_0|)+p(\lambda|X_2Y_1)(|m_2+n_1-1|)]\\
		\leq&4+4\alpha-36(2G-1)\sum_\lambda p(\lambda)\min p(X_j, Y_k|\lambda)-18(\alpha-1)\sum_\lambda p(\lambda)\min p(X_j, Y_k|\lambda)\\&|m_0+m_1-n_0-n_1|\\
		\leq&4+4\alpha-36(2G-1)\sum_\lambda p(\lambda)\min\frac{p^{'}(X_j, Y_k|\lambda)-S}{1-9S}-18(\alpha-1)\sum_\lambda p(\lambda)\min\frac{p^{'}(X_j, Y_k|\lambda)-S}{1-9S}\\&|m_0+m_1-n_0-n_1|\\
		\leq&(1-9S)[4+4\alpha-36(2G-1)\sum_\lambda p^{'}(\lambda)\min(p(X_j, Y_k|\lambda)-S)]-18(\alpha-1)\sum_\lambda p(\lambda)\\&\min (p^{'}(X_j, Y_k|\lambda)-S)|m_0+m_1-n_0-n_1|.
	\end{aligned}
\end{equation}

According to equations (\ref{E15}) and (\ref{E16}), we obtain that the relationship between $ I^\alpha_{3322} $ and $ \overline{I}^\alpha_{3322} $ holds the following form
\begin{equation}
	\label{E17}
	I^\alpha_{3322}(G,S,P)=(1-9S)\overline{I}^\alpha_{3322}(G,P)+36(\alpha -1)S+36S.
\end{equation}
Therefore, we give the following conclusions:\\
(1) When $ \dfrac{P-S}{1-9S}\ge \dfrac{1}{8} $, i.e., $ 8P+S\ge 1 $, we have
\begin{equation}
	\begin{aligned}
		\label{E18}
		I^\alpha_{3322}(G,S,P)&=(1-9S)\overline{I}^\alpha_{3322}(G,P)+36(\alpha -1)S+36S\\&=4+4\alpha-36S.
	\end{aligned}
\end{equation}
\\(2) When $ \dfrac{P-S}{1-9S}< \dfrac{1}{8} $, i.e., $ 8P+S< 1 $, we have
\begin{equation}
	\begin{aligned}
		\label{E19}
		I^\alpha_{3322}(G,S,P)&=(1-9S)\overline{I}^\alpha_{3322}(G,P)+36(\alpha -1)S+36S\\&=\overline{I}^\alpha_{3322}(G,P).
	\end{aligned}
\end{equation}
To summarize the above derivation, we get
\begin{equation}
	\label{E71}
	\\I^\alpha_{3322}(G,S,P)=
	\begin{cases}
		4\alpha + 4 - 36S,&\text{$8P + S \geq 1$},\\
		4\alpha + 4 - 36(2G - 1)(1 - 8P),&
		\text{$8P + S < 1$}.
	\end{cases}\
\end{equation}
Here we complete the proof of Theorem \ref{th1}. 
\qed

Based on Theorem \ref{th1}, we take DI randomness expansion as an example and use deterministic strategy (i.e., $ G = 1 $)  to analyze the results under different cases. 

\emph{Case 1.} $ P = \dfrac{1}{9} $ or $ S = \dfrac{1}{9} $. It means that Eve can't get any information about input with the local hidden variables $\lambda$. Eve attempts to forge inequality violations by using predefined strings as the output of the device (i.e., $ G = 1 $). But from Theorem \ref{th1}, we can find that the maximum value $ I^\alpha_{3322}(1,S,P) $ will not exceed $ 4\alpha $ at this time, so Eve cannot successfully fake the violation of 1-PFB inequality.
\begin{figure}[H]
	\centering     
	\subfigure[]{                    
		\begin{minipage}{7cm}
			\centering  
			\includegraphics[scale=0.4]{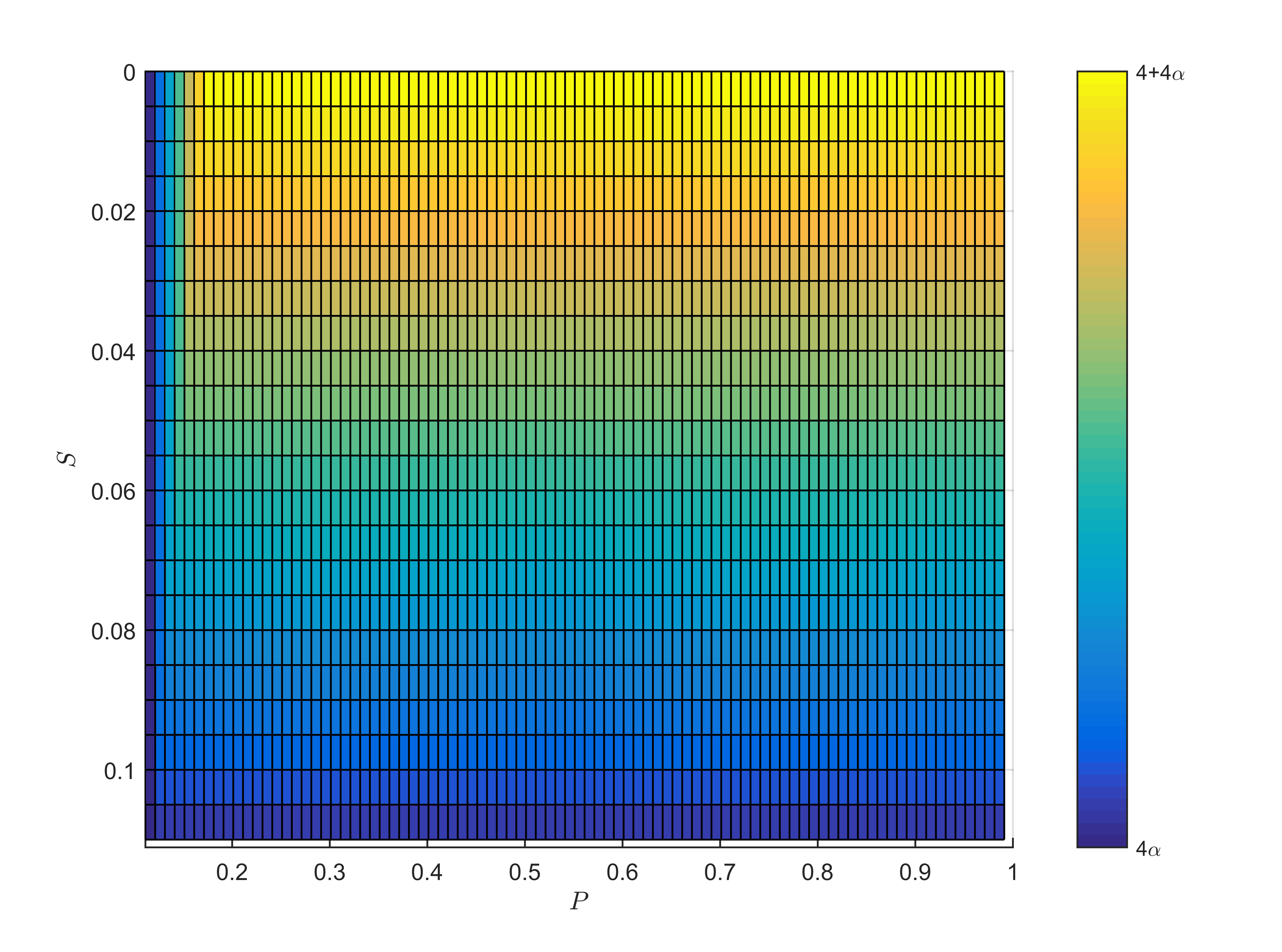} 
			\label{fig1(a)}
	\end{minipage}}
	\subfigure[]{                   
		\begin{minipage}{7cm}
			\centering  
			\includegraphics[scale=0.4]{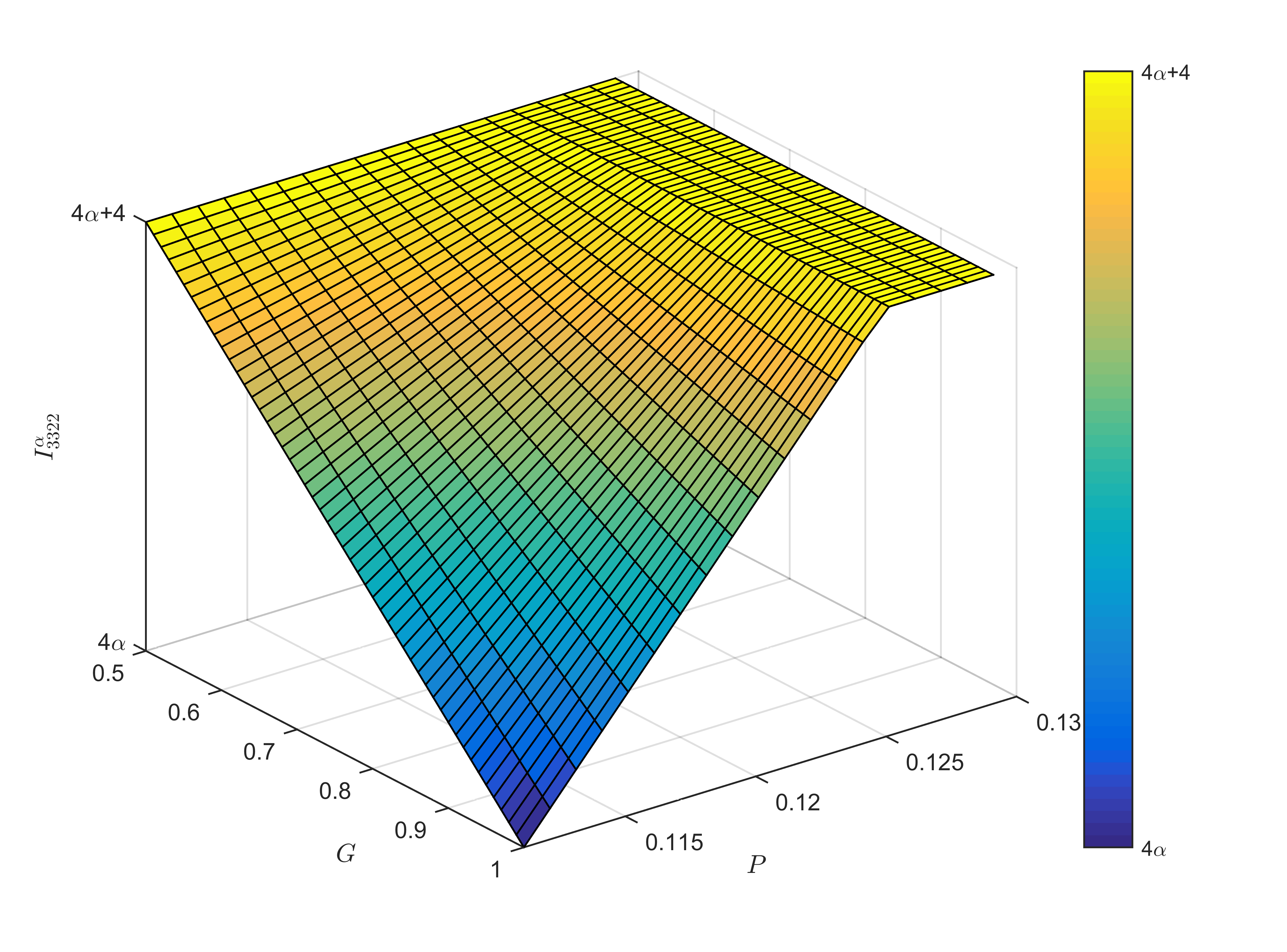}       
			\label{fig1(b)}
	\end{minipage}}
	\caption{Schematic of $ I^\alpha_{3322}(1,S,P) $ and  $ I^\alpha_{3322}(G,P) $ under the general input distribution. (a) The maximum value of 1-PFB correlation function that Eve can fake, $ I^\alpha_{3322}(1,S,P) $, is plotted against measurement dependence $ P $, $ S $ and completely deterministic strategy (i.e., $ G = 1 $ ) under the general input distribution. And the color depth indicates the size of $ I^\alpha_{3322}(1,S,P) $. (b) The maximum value of 1-PFB correlation function that Eve can fake,  $ I^\alpha_{3322}(G,P) $, is plotted against measurement dependence $ P $ and guessing probability $ G $ under the general input distribution. }  
	\label{fig1}  
\end{figure}

\emph{Case 2.} $ 8P + S > 1 $ or $ 8P + S < 1 $. It means that Eve can get some information about input through the local hidden variables $\lambda$. In this case, Eve attempts to fake inequality violations by using predefined strings as the output of the device (i.e., $ G = 1 $). We can find $ 4\alpha \leq I^\alpha_{3322}(1,S,P) \leq 4\alpha + 4 $ from Fig. \ref{fig1(a)}, so Eve can use a deterministic strategy to successfully fake the violation of 1-PFB inequality.

Obviously, true randomness cannot be generated in \emph{Case 1}, so we are interested in whether true randomness is generated in \emph{Case 2}. For the sake of more obvious conclusion, we show the relationship between $ P $ and $ I^\alpha_{3322} $ when guessing probability takes different values in Fig. \ref{fig1(b)}. We find that as long as the observed value satisfies $ I^\alpha_{3322,obv} > I^\alpha_{3322}(1,S,P) $, we can ensure that true randomness is generated in the process.
Then we will verify the compactness of equation (\ref{E7}) given in Theorem 1 by giving an optimal strategy. In deterministic theory, we admit that
\begin{equation}
	\begin{split}
		\label{30}
		p(a|x,\lambda) = \delta_{a,a_\lambda(x)},\\
		p(b|y,\lambda) = \delta_{b,b_\lambda(y)},
	\end{split}
\end{equation}
where $ a_\lambda(x) $ and $ b_\lambda(y) $ indicate the output under a certain deterministic strategy. Therefore we get an expression for $ I^\alpha_{3322} $
\begin{equation}
	\begin{split}
		\label{31}
		I^\alpha_{3322} =& 9\sum_\lambda[p(\lambda)p(X_0, Y_0|\lambda)a_\lambda(0)b_\lambda(0)+p(\lambda)p(X_0, Y_1|\lambda)a_\lambda(0)b_\lambda(1)\\
		&+p(\lambda)p(X_1, Y_0|\lambda)a_\lambda(1)b_\lambda(0)+p(\lambda)p(X_1, Y_1|\lambda)a_\lambda(1)b_\lambda(1)]\\
		&+9a\sum_\lambda[p(\lambda)p(X_0, Y_2|\lambda)a_\lambda(0)b_\lambda(2)-p(\lambda)p(X_1, Y_2|\lambda)a_\lambda(1)b_\lambda(2)\\
		&+p(\lambda)p(X_2, Y_0|\lambda)a_\lambda(2)b_\lambda(0)-p(\lambda)p(X_2, Y_1|\lambda)a_\lambda(2)b_\lambda(1)].
	\end{split}
\end{equation}
\begin{table}[!htbp]
	\centering
	\setlength{\abovecaptionskip}{0pt}
	\setlength{\belowcaptionskip}{10pt}
	\caption{The output is determined by $\lambda$, $ j $ and $ k $.}\label{tab:aStrangeTable1}
	\begin{tabular}{p{2.3cm}p{2.3cm}p{2.3cm}p{2.3cm}p{2.3cm}p{2.3cm}lllllll}
		\toprule[1pt]
		$\lambda_n$ & $ a_0 $ & $ b_0 $ & $ a_1 $ & $ b_1 $ & $ a_2 $ & $ b_2 $\\
		\midrule
		$\lambda_0$ &   -1   & 1 & 1 & -1 & 1 & -1\\
		\midrule
		$\lambda_1$ &   1   & 1 & -1 & -1 & 1 & 1\\
		\midrule
		$\lambda_2$ &   1   & -1 & -1 & 1 & -1 & 1\\
		\midrule
		$\lambda_3$ &   1   & 1 & -1 & -1 & 1 & 1\\
		\bottomrule[1pt]
	\end{tabular}
	\label{tab1}
\end{table}
Then, we give the optimal strategy corresponding to the maximum value of 1-PFB correlation function that Eve can fake (see Table \ref{tab1}). Based on the strategy in Table \ref{tab1}, we can simplify equation (\ref{31}) to
\begin{equation}
	\begin{aligned}
		\label{32}
		I^\alpha_{3322} =& 4 + 4\alpha - \dfrac{9}{2}(p(X_0, Y_0|\lambda_0) + p(X_0, Y_0|\lambda_2)+p(X_0, Y_1|\lambda_1) + p(X_0, Y_1|\lambda_3)\\&+p(X_1, Y_0|\lambda_1) + p(X_1, Y_0|\lambda_3)+p(X_1, Y_1|\lambda_0) + p(X_1, Y_1|\lambda_2)),
	\end{aligned}
\end{equation}
where equation (\ref{32}) is based on $ p(\lambda_n)=\dfrac{1}{4} $, $ P(X_j, Y_k|\lambda) = \dfrac{1}{9} $ and $ \displaystyle\sum_\lambda{p(\lambda)p(X_j, Y_k|\lambda)} = p(X_j, Y_k) $. Next, we consider the value of $ I^\alpha_{3322} $ in different cases:\\
(1) When $ 8P + S \geq 1 $, let $ p(X_0, Y_0|\lambda_0) = p(X_0, Y_0|\lambda_2)=p(X_0, Y_1|\lambda_1)=p(X_0, Y_1|\lambda_3)=p(X_1, Y_0|\lambda_1)=p(X_1, Y_0|\lambda_3)=p(X_1, Y_1|\lambda_0) = p(X_1, Y_1|\lambda_2)=S $, we have $ \max I^\alpha_{3322} = 4 + 4\alpha -36S $.\\
(2) When $ 8P + S < 1 $, we have
\begin{equation}
	\begin{aligned}
		\label{322}
		I^\alpha_{3322} =& 4 + 4\alpha - \dfrac{9}{2}(p(X_0, Y_0|\lambda_0) + p(X_0, Y_0|\lambda_2)+p(X_0, Y_1|\lambda_1) + p(X_0, Y_1|\lambda_3)+p(X_1, Y_0|\lambda_1) \\&+ p(X_1, Y_0|\lambda_3)+p(X_1, Y_1|\lambda_0) + p(X_1, Y_1|\lambda_2))\\=&4 + 4\alpha - \dfrac{9}{2}(1-\displaystyle\sum_{(j,k)\in\{0,1,2\}^2/\{0,0\}}p(X_j, Y_k|\lambda_0)+1-\displaystyle\sum_{(j,k)\in\{0,1,2\}^2/\{0,0\}}p(X_j, Y_k|\lambda_2)\\&+1-\displaystyle\sum_{(j,k)\in\{0,1,2\}^2/\{0,1\}}p(X_j, Y_k|\lambda_1)+1-\displaystyle\sum_{(j,k)\in\{0,1,2\}^2/\{0,1\}}p(X_j, Y_k|\lambda_3)\\&+1-\displaystyle\sum_{(j,k)\in\{0,1,2\}^2/\{1,0\}}p(X_j, Y_k|\lambda_1)+1-\displaystyle\sum_{(j,k)\in\{0,1,2\}^2/\{1,0\}}p(X_j, Y_k|\lambda_3)\\&+1-\displaystyle\sum_{(j,k)\in\{0,1,2\}^2/\{1,1\}}p(X_j, Y_k|\lambda_0)+1-\displaystyle\sum_{(j,k)\in\{0,1,2\}^2/\{1,1\}}p(X_j, Y_k|\lambda_2)),   
	\end{aligned}
\end{equation}
let $ p(X_j, Y_k|\lambda)=P $ except for $ p(X_0, Y_0|\lambda_0) $, $ p(X_0, Y_0|\lambda_2) $, $ p(X_0, Y_1|\lambda_1) $, $ p(X_0, Y_1|\lambda_3) $, $ p(X_1, Y_0|\lambda_1) $, $ p(X_1, Y_0|\lambda_3) $, $ p(X_1, Y_1|\lambda_0) $ and $ p(X_1, Y_1|\lambda_2) $, we get $ \max I^\alpha_{3322} = 4 + 4\alpha -36(1-8P) $.

Evidently, it satisfies the maximum value of $ I^\alpha_{3322} $ we obtained previous. So far, we have found a deterministic strategy that enables Eve to fake maximum value of $ I^\alpha_{3322} $.
In the latter section, we need to compare the difficulty of the maximum quantum violation of 1-PFB inequality and Chain inequality Eve forged by using the relationship between the maximum value of Chain correlation function that Eve can fake and measurement dependence. Using the method shown in the derivation process of Theorem \ref{th1}, it is easy to get the maximum value of the Chain correlation function faked by Eve. So we no longer give detailed analysis, proof and optimal strategy here, but only give the relationship between the maximum value of the Chain correlation function that Eve can fake under general input distribution in Theorem \ref{th2}.
\begin{theorem}
	\label{th2}
	The maximum value of Chain correlation function with three measurement settings that Eve can fake, $ I_{Chain}(G,S,P) $, for any no-signaling model with $ p(X_j,Y_k) = \dfrac{1}{9} $, is
	\begin{equation}
		\label{E71}
		\\I_{Chain}(G,S,P)=
		\begin{cases}
			6 - 36S,&\text{$8P + S \geq 1$},\\
			6 - 18(2G - 1)(1 - 8P),&
			\text{$8P + S < 1$},
		\end{cases}\
	\end{equation}
	where the definitions of $ G $, $ S $ and $ P $ are the same as before.
\end{theorem}
\subsection{The factorizable input distribution}
\label{sec32}
Next, we give the relationship among measurement dependence, guessing probability, and 
the maximum value of 1-PFB correlation function that Eve can fake under the factorizable input distribution.
\begin{theorem}
	\label{th3}
	The maximum value of 1-PFB correlation function that Eve can fake, $ I^\alpha_{3322}(G,S,P) $, for any no-signaling model with $ p(X_j,Y_k) = \dfrac{1}{9} $, is
	\begin{equation}
		\label{33}
		\\I^\alpha_{3322}(G,S,P)=
		\begin{cases}
			4\alpha + 4 - 36S,&\text{$2P + S \geq \dfrac{1}{3}$},\\
			4\alpha + 4 - 12(2G - 1)(1 - 6P),&
			\text{$2P + S < \dfrac{1}{3}$},
		\end{cases}\
	\end{equation}
	where the definitions of $ G $, $ S $ and $ P $ are the same as before.
\end{theorem}
\indent \textbf{{Proof}}
Similar to the proof of Theorem \ref{th1}, we first consider the case of a fixed lower bound of measurement dependence. Let $ P=P_{A}P_{B} $, where $ P_{A}=\max P_{A}(X_{j}|\lambda) $ and $ P_{B}=\max P_{B}(Y_{k}|\lambda) $. We have
\begin{equation}
	\begin{split}
		\label{34}
		\min P(X_{j},Y_{k}|\lambda)&=\min P_{A}(X_{j}|\lambda)\min P_{B}(Y_{k}|\lambda)\\&=1-2(P_{A}+P_{B})+4P.
	\end{split}
\end{equation}
Based on equation (\ref{E14}), we analyze the value of $ 	\min P(X_{j},Y_{k}|\lambda) $ for the factorizable input distribution as follows:\\
(1) Suppose that $ P\ge \dfrac{1}{6} $, we always discover that $ \min P(X_{j},Y_{k}|\lambda)=0 $.\\
(2) Suppose that $ \dfrac{1}{9}\le P\le \dfrac{1}{6} $, we find that  $ \min P(X_{j},Y_{k}|\lambda)=\dfrac{1}{3}-2P $.

So, 1-PFB inequality for the fixed lower bound of the measurement dependence can be obtained by
\begin{equation}
	\label{35}
	\\\overline{I}^\alpha_{3322}(G,P)=
	\begin{cases}
		4\alpha + 4,&\text{$P\geq \dfrac{1}{6}$},\\
		4\alpha + 4 - 12(2G - 1)(1 - 6P),&
		\text{$\dfrac{1}{9}\le P < \dfrac{1}{6}$}.
	\end{cases}\
\end{equation}
Similarly, we can also obtain the case of flexible lower bound 
\begin{equation}
	\label{36}
	\\I^\alpha_{3322}(G,S,P)=
	\begin{cases}
		4\alpha + 4 -36S,&\text{$2P+S\geq \dfrac{1}{3}$},\\
		4\alpha + 4 - 12(2G - 1)(1 - 6P),&
		\text{$2P+S< \dfrac{1}{3}$}.
	\end{cases}\
\end{equation}
The proof of Theorem \ref{th3} is completed.
\qed

Here, we only analyze the circumstance in which the output contains true randomness, and the analysis in other cases is similar to Theorem \ref{th1}. From Theorem \ref{th3}, we find that when the factorizable input satisfies $ 2P + S > \dfrac{1}{3} $ or $ 2P + S < \dfrac{1}{3} $, Eve can forge the violation of 1-PFB inequality. The maximum value of 1-PFB correlation function is given in Fig. \ref{fig2(a)}. We find in Fig. \ref{fig2(b)} that when the observation value satisfies $ I^\alpha_{3322,obv} > I^\alpha_{3322}(1,S,P) $, the output contains true randomness.
\begin{figure}[H]
	\centering     
	\subfigure[]{                    
		\begin{minipage}{7cm}
			\centering  
			\includegraphics[scale=0.4]{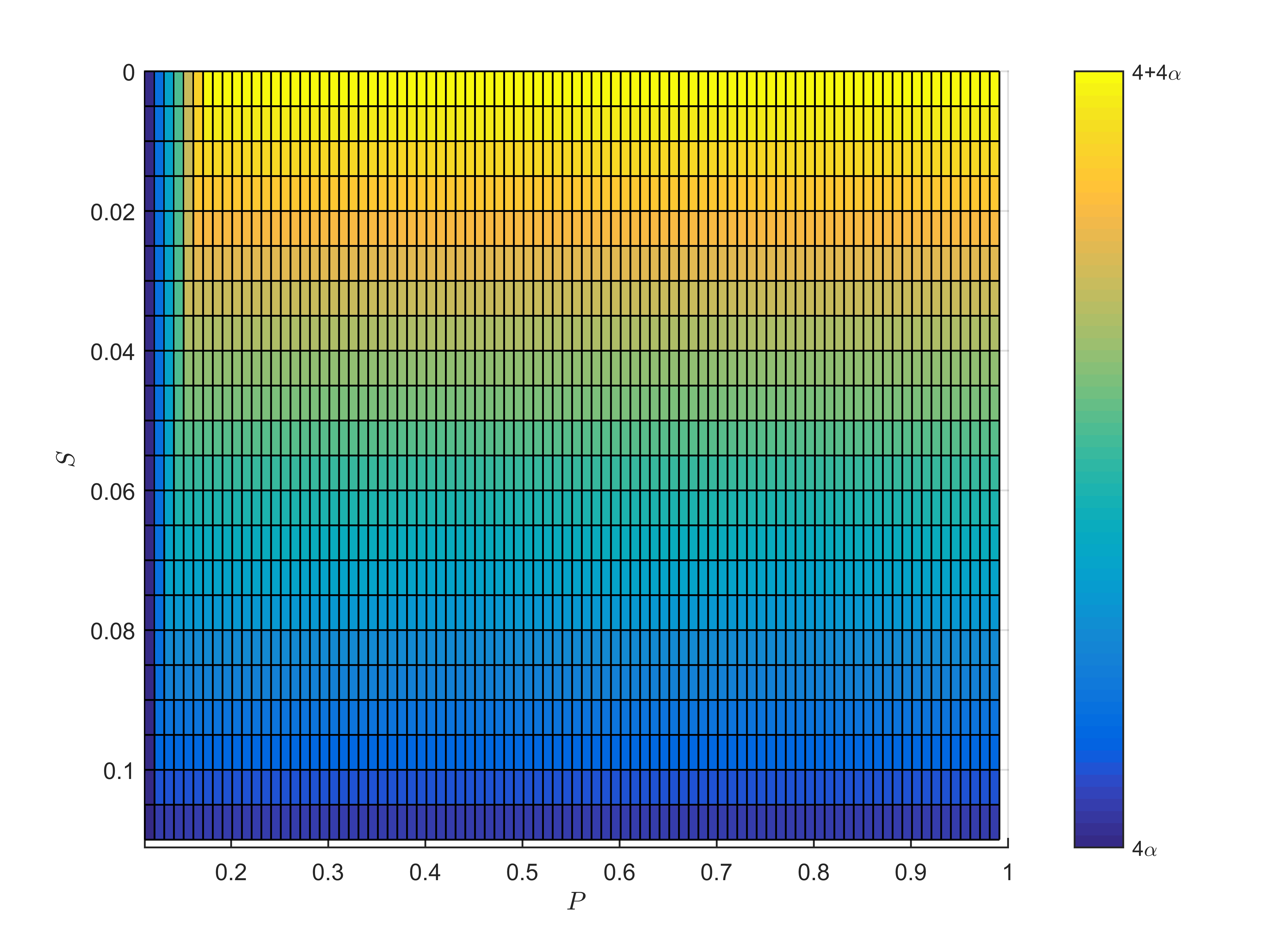} 
			\label{fig2(a)}
	\end{minipage}}
	\subfigure[]{                   
		\begin{minipage}{7cm}
			\centering  
			\includegraphics[scale=0.4]{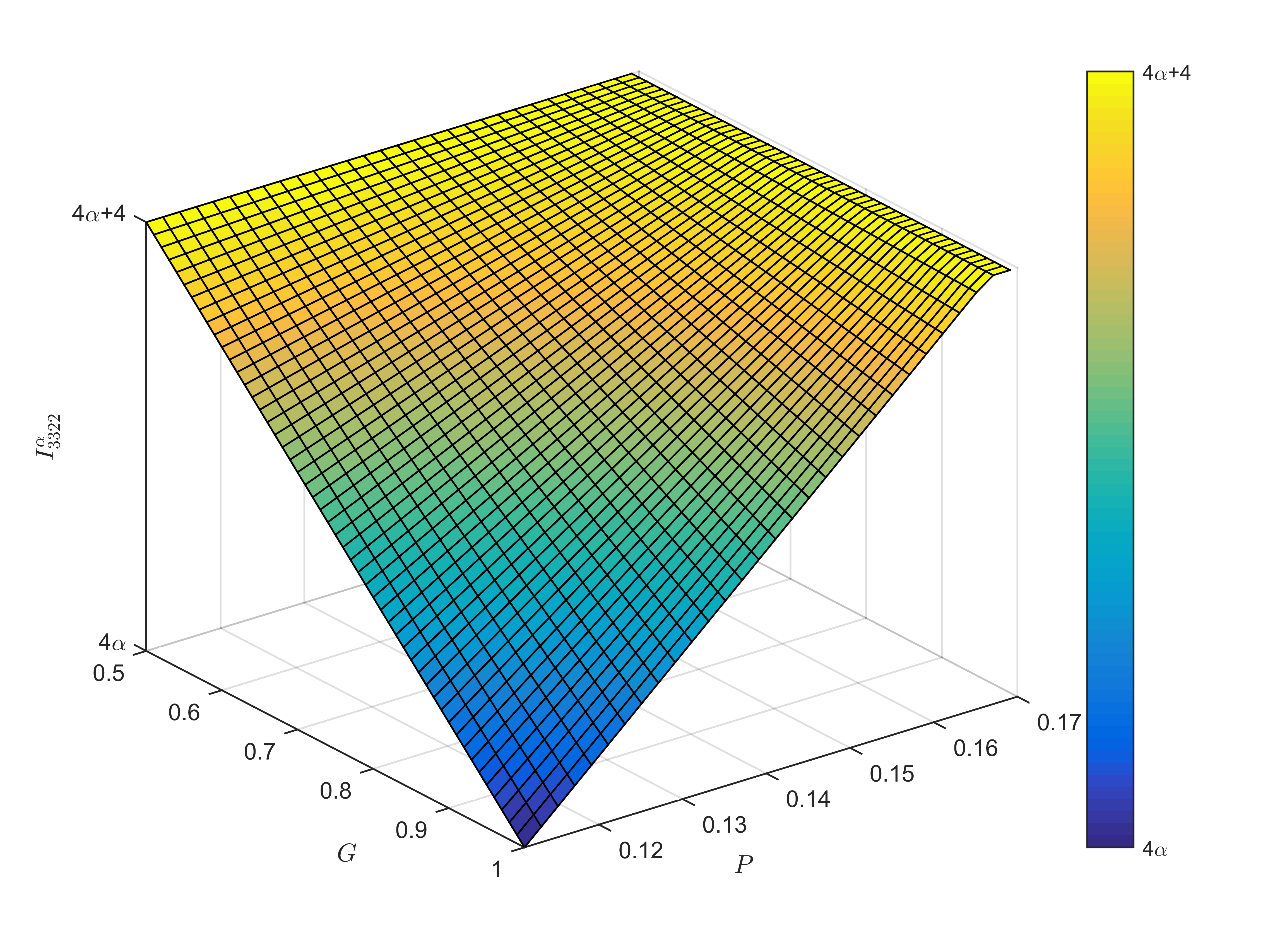}       
			\label{fig2(b)}
	\end{minipage}}
	\caption{Schematic of $ I^\alpha_{3322}(1,S,P) $ and  $ I^\alpha_{3322}(G,P) $ under the factorizable input distribution. (a) The maximum value of 1-PFB correlation function that Eve can fake, $ I^\alpha_{3322}(1,S,P) $, is plotted against measurement dependence $ P $, $ S $ and completely deterministic strategy (i.e., $ G = 1 $ ) under the factorizable input distribution. And the color depth indicates the size of $ I^\alpha_{3322}(1,S,P) $. (b) The maximum value of 1-PFB correlation function that Eve can fake,  $ I^\alpha_{3322}(G,P) $, is plotted against measurement dependence $ P $ and guessing probability $ G $ with the factorizable input distribution. }  
	\label{fig1}  
\end{figure}

Similar to the analysis under the general input distribution, we give the optimal strategy of 1-PFB correlation function under the factorizable input distribution
\begin{equation}
	\begin{split}
		\label{34}
		I^\alpha_{3322} =& 9\sum_\lambda{p(\lambda)}\sum_{a,b}[p(X_0, Y_0|\lambda)p(a|X_0,\lambda)p(b|Y_0,\lambda)-p(X_0, Y_1|\lambda)p(a|X_0,\lambda)p(b|Y_1,\lambda)\\
		&-p(X_1, Y_0|\lambda)p(a|X_1,\lambda)p(b|Y_0,\lambda)+p(X_1, Y_1|\lambda)p(a|X_1,\lambda)p(b|Y_1,\lambda)]\\
		&+9\alpha\sum_\lambda{p(\lambda)}\sum_{a,b}[p(X_0, Y_2|\lambda)p(a|X_0,\lambda)p(b|Y_2,\lambda)-p(X_1, Y_2|\lambda)p(a|X_1,\lambda)p(b|Y_2,\lambda)\\
		&+p(X_2, Y_0|\lambda)p(a|X_2,\lambda)p(b|Y_0,\lambda)-p(X_2, Y_1|\lambda)p(a|X_2,\lambda)p(b|Y_1,\lambda)].
	\end{split}
\end{equation}

Using the strategy in Ref. \cite{QRNG10} for a given local hidden variable and an arbitrary input, we have
\begin{equation}
	\begin{split}
		\label{35}
		p(0,x|\lambda) = p(0,y|\lambda) = 1,\\
		p(1,x|\lambda) = p(1,y|\lambda) = 0.
	\end{split}
\end{equation}
It is easy to verify that this strategy can satisfy the maximum value of $ I^\alpha_{3322}(G,S,P) $ we gave in Theorem \ref{th3}. 
Similar to Theorem \ref{th2}, we give the relationship among measurement dependence, guessing probability and the maximum value of Chain correlation function that Eve can forge under the factorizable input distribution in Theorem \ref{th4}.
\begin{theorem}
	\label{th4}
	The maximum value of Chain correlation function with three measurement settings that Eve can fake, $ I_{Chain}(G,S,P) $, for any no-signaling model with $ p(X_j,Y_k) = \dfrac{1}{9} $, is
	\begin{equation}
		\label{E36}
		\\I_{Chain}(G,S,P)=
		\begin{cases}
			6 - 36S,&\text{$2P + S \geq \dfrac{1}{3}$},\\
			6 - 6(2G - 1)(1 - 6P),&
			\text{$2P + S < \dfrac{1}{3}$},
		\end{cases}\
	\end{equation}
	where the definitions of $ G $, $ S $ and $ P $ are the same as before.
\end{theorem}
\subsection{Comparison between 1-PFB inequality and Chain inequality}
\label{sec33}
In subsection \ref{sec31} and \ref{sec32}, we establish the relationship among measurement dependence, guessing probability, and the maximum value of 1-PFB inequality and Chain correlation function that Eve can fake under different input distributions. 
\begin{table}[!htbp]
	\centering
	\setlength{\abovecaptionskip}{0pt}
	\setlength{\belowcaptionskip}{10pt}
	\caption{We compare the unknown information rate of the 1-PFB inequality and the  Chain inequality when Eve forges the  respective maximum quantum violation under deterministic strategy.}\label{tab:aStrangeTable2}
	\begin{tabular}{p{5.5cm}p{5.5cm}p{5.5cm}lll}
		\toprule
		The type of inequality & General input distribution &  Factorizable input distribution\\
		\midrule
		Chain inequality(three measurement settings) &    0.969   & 0.882\\
		\midrule
		1-PFB inequality &   $-\dfrac{1}{2}\log_{3}{\dfrac{ 36-4\alpha +\alpha^2 }{288}}$  & $-\dfrac{1}{2}\log_{3}{\dfrac{ 12 -4\alpha+\alpha^2}{72}}$\\
		\bottomrule
	\end{tabular}
	\label{tab2}
\end{table}

We are concerned about whether it is more difficult for Eve to fake the violation in 1-PFB inequality than in Chain inequality. To get the answer, we compare the critical value of measurement dependence that Eve uses on deterministic strategy to fake the maximum quantum violation (i.e., $ 4+\alpha^2 $ and $ 3\sqrt{3} $, respectively) in different inequalities. Here, in order to make the conclusion clearer, we use the unknown information rate  $ \tau_M=-\dfrac{1}{2}\log_{M}\hat{P} $ defined in Ref. \cite{QRNG12}, where $ M $ represents the number of measurement settings (in this paper, $ M=3 $) and $ \hat{P} $ represents the degree of measurement dependence when Eve reaches the maximum quantum violation. From the expression of unknown information rate $ \tau_M $ that the larger $ \tau_M $ is, the closer $ \hat{P} $ is to $ \dfrac{1}{M^2} $ (i.e., Eve exploits less information
about inputs to fake the maximum quantum violation). For the comparison of unknown information rate $ \tau_M $ of two inequalities, we show them in Table \ref{tab2}. In order to compare the unknown information rate of the two inequalities more intuitively, we show the change of the unknown information rate under different input distributions in Fig. 5. 
\begin{figure}[H]
	\centering     
	\subfigure[]{                    
		\begin{minipage}{7cm}
			\centering  
			\includegraphics[scale=0.4]{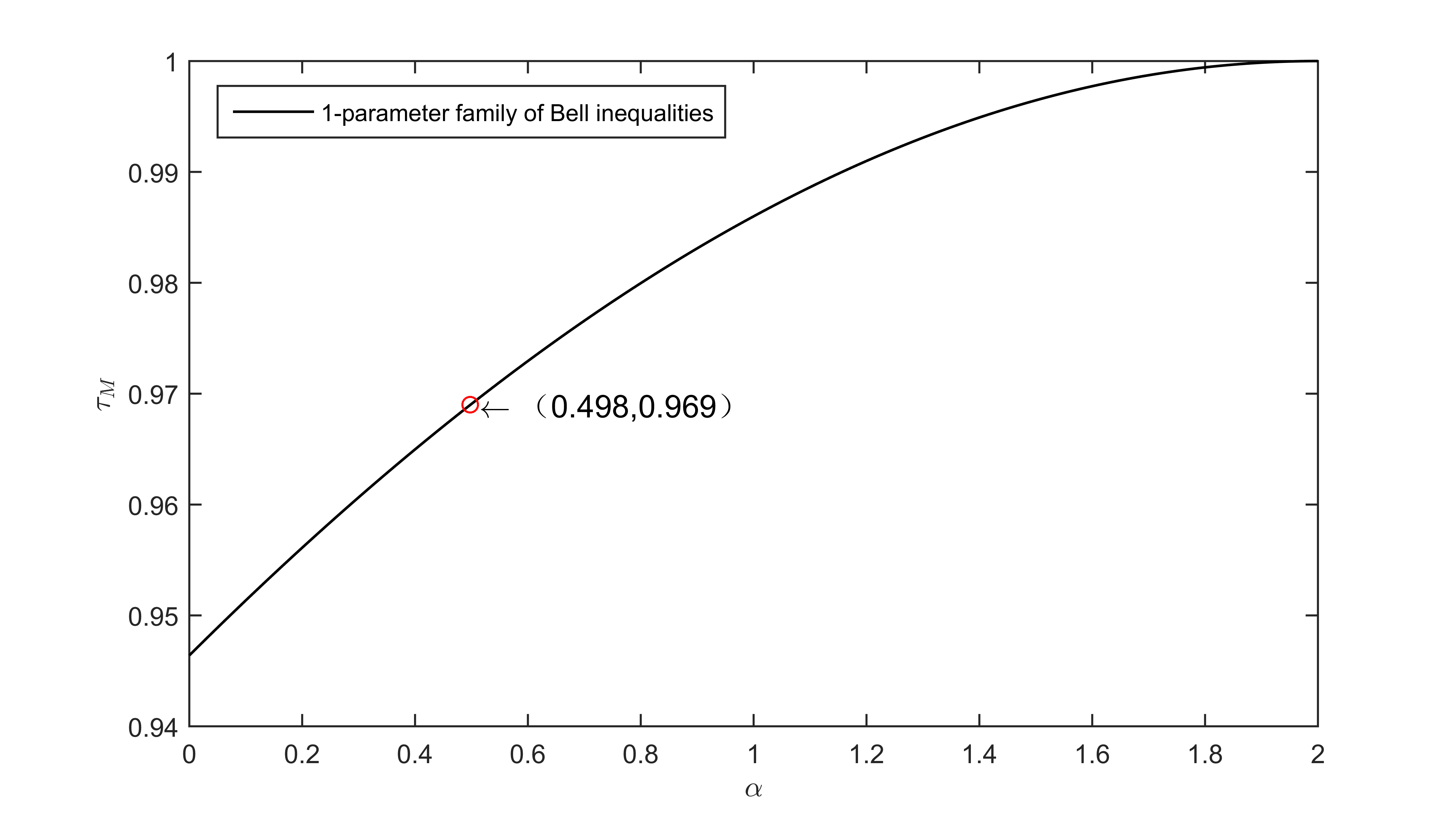} 
	\end{minipage}}
	\subfigure[]{                   
		\begin{minipage}{7cm}
			\centering  
			\includegraphics[scale=0.4]{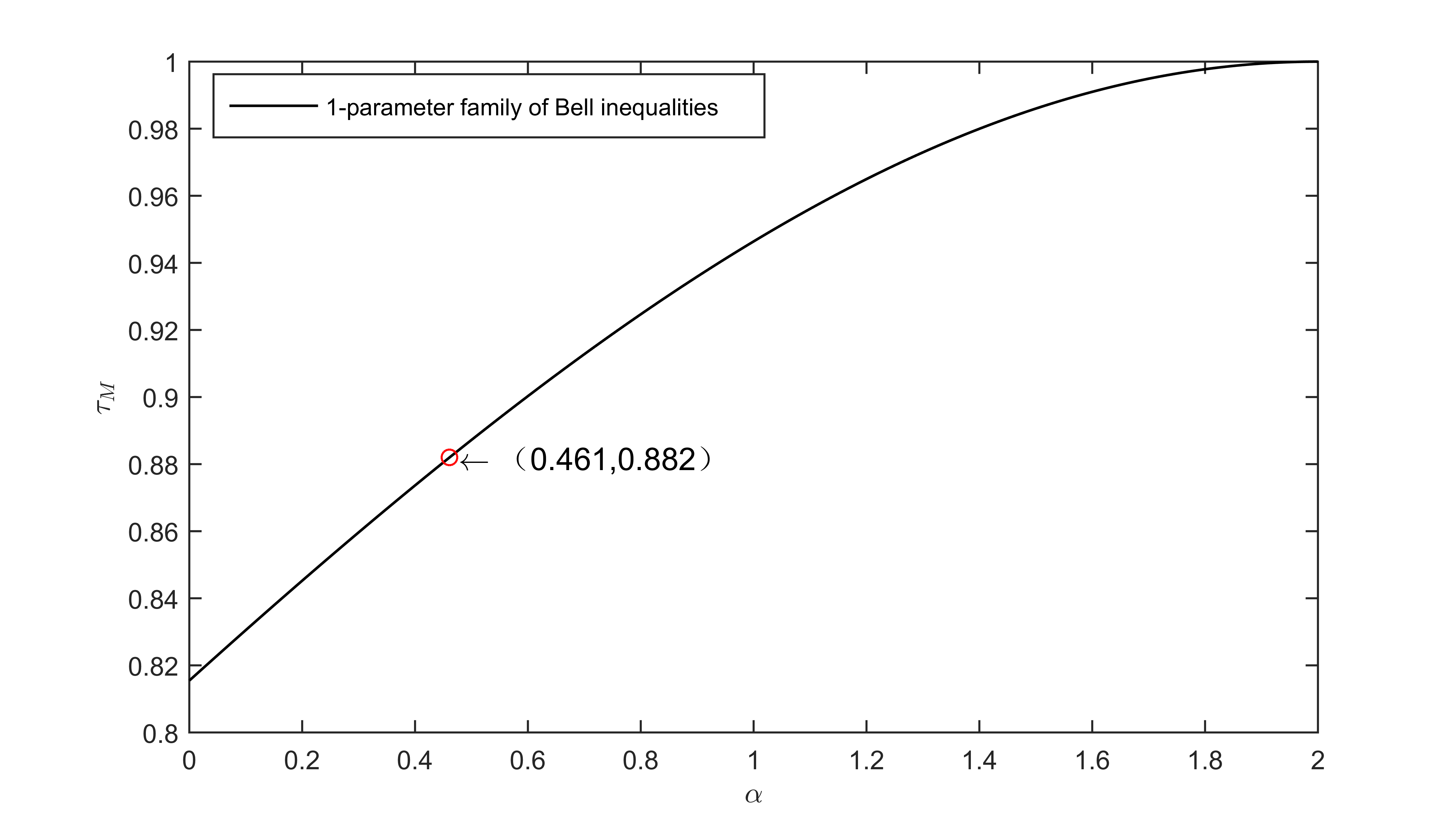}       
	\end{minipage}}
	\caption{The critical value of unknown information rate $ \tau_M $ of 1-PFB inequality under different input distributions. The red dots represent the value of $ \alpha $ when 1-PFB inequality and Chain inequality take the same unknown information rate $ \tau_M $. (a) The critical value of unknown information rate $ \tau_M $ of 1-PFB inequality under general input distributions. (b) The critical value of unknown information rate $ \tau_M $ of 1-PFB inequality under factorizable input distributions.}  
	\label{fig5}  
\end{figure}
According to Fig. \ref{fig5}, we draw two conclusions:\\
(1) Under the general input distribution, it is more difficult for Eve to fake the maximum quantum violation in 1-PFB inequality than in Chain inequality when $ \alpha<0.498 $, and in other cases, just the opposite.\\
(2) Under the factorizable input distribution, it is more difficult for Eve to fake the maximum quantum violation in 1-PFB inequality than in Chain inequality when $ \alpha<0.461 $, and in other cases, we come to the opposite conclusion.\\
\section{\textbf{Conclusions}}
\label{sec4}
Measurement independence is one of the assumptions based on which Bell tests can detect quantum nonlocality in DI scenario. In the DI framework where each party has three measurement settings and two outputs, we studied the effects of relaxing the assumption of measurement independent on 1-PFB tests. For both general and factorizable input distributions, we established the relationship among measurement dependence involving in flexible upper and lower bound, guessing probability, and the maximum value of 1-PFB correlation function that Eve can fake. At the same time, we gave a deterministic strategy when Eve forged the maximum value of 1-PFB correlation function. In addition, we found that when the flexible upper and lower bounds of measurement dependence satisfy certain conditions, the output may contain true randomness. We also explored the effects of measurement dependence with flexible 
upper and lower bound on Chain inequality.  Moreover, 
by comparing the unknown information rate of Chain inequality and 1-PFB inequality, we gave the range of parameter $ \alpha $ when it is more difficult for Eve to forge the maximum quantum violation in 1-PFB inequality than in Chain inequality. It's interesting to study whether each party's input selection can be controlled by a different Eve, and we expect our derivation method to provide ideas for the study of this problem.
\section{\textbf{ACKNOWLEDGMENTS}}
	This work is supported by NSFC (Grant Nos. 61973021, 61672110, 61671082, 61976024, 61972048), and the Fundamental Research Funds for the Central Universities (Grant No.2019XD-A01).


\begin{thebibliography}{}
	\bibitem{quantum nonlocality1}
	Popescu S, Rohrlich D.:Quantum nonlocality as an axiom[J]. Foundations of Physics, 1994, 24(3): 379-385.
	\bibitem{quantum nonlocality2}
	Barrett J, Linden N, Massar S, et al.:Nonlocal correlations as an information-theoretic resource[J]. Physical Review A, 2005, 71(2): 022101.
	\bibitem{quantum nonlocality3}
	Gallego R, Würflinger L E, Acín A, et al.:Quantum correlations require multipartite information principles[J]. Physical review letters, 2011, 107(21): 210403.
	\bibitem{quantum nonlocality4}
	Christensen B G, Liang Y C, Brunner N, et al.:Exploring the limits of quantum nonlocality with entangled photons[J]. Physical Review X, 2015, 5(4): 041052.
	\bibitem{QKD1}
	Ekert A K.:Quantum cryptography based on Bell’s theorem[J]. Physical review letters, 1991, 67(6): 661.
	\bibitem{QKD2}
	Acín A, Gisin N, Masanes L.:From Bell’s theorem to secure quantum key distribution[J]. Physical review letters, 2006, 97(12): 120405.
	\bibitem{robust}
	Bouda J, Pawłowski M, Pivoluska M, et al.:Device-independent randomness extraction from an arbitrarily weak min-entropy source[J]. Physical Review A, 2014, 90(3): 032313.
	\bibitem{QRNG1}
	Pironio S, Acín A, Massar S, et al.:Random numbers certified by Bell’s theorem[J]. Nature, 2010, 464(7291): 1021-1024.
	\bibitem{QRNG2}
	Pironio S, Massar S.:Security of practical private randomness generation[J]. Physical Review A, 2013, 87(1): 012336.
	\bibitem{entanglement certification1}
	Bancal J D.:Device-independent witnesses of genuine multipartite entanglement[M]//On the Device-Independent Approach to Quantum Physics. Springer, Cham, 2014: 73-80.
	\bibitem{entanglement certification2}
	Barreiro J T, Bancal J D, Schindler P, et al.:Demonstration of genuine multipartite entanglement with device-independent witnesses[J]. Nature Physics, 2013, 9(9): 559-562.
	\bibitem{Bell test1}
	Bell J S.:On the einstein podolsky rosen paradox[J]. Physics Physique Fizika, 1964, 1(3): 195.
	\bibitem{Bell test2}
	Bell J S, Aspect A.:Speakable and Unspeakable in Quantum Mechanics: How to teach special relativity[J]. Cambridge University Press, 2004, 57(6):567-567.
	\bibitem{Bell test3}
	Clauser J F, Horne M A, Shimony A, et al.:Proposed Experiment to Test Local Hidden Variable Theories[J]. Physical Review Letters, 1970, 24(10): 549.
	\bibitem{Bell test4}
	Wehner S.:Tsirelson bounds for generalized clauser-horne-shimony-holt inequalities[J]. Physical Review A, 2006, 73(2): 022110.
	\bibitem{QRNG4}
	Hall M J W.:Relaxed bell inequalities and kochen-specker theorems[J]. Physical Review A, 2011, 84(2): 022102.
	\bibitem{QRNG5}
	Hall M J W.:Local deterministic model of singlet state correlations based on relaxing measurement independence[J]. Physical review letters, 2010, 105(25): 250404.
	\bibitem{QRNG6}
	Koh D E, Hall M J W, Pope J E, et al.:Effects of reduced measurement independence on Bell-based randomness expansion[J]. Physical review letters, 2012, 109(16): 160404.
	\bibitem{QRNG7}
	Sheridan L, Scarani V.:Bell tests with min-entropy sources[J]. Physical Review A, 2013, 87(6): 062121.
	\bibitem{QRNG8}
	Pope J E, Kay A.:Limited measurement dependence in multiple runs of a Bell test[J]. Physical Review A, 2013, 88(3): 032110.
	\bibitem{QRNG9}
	Pütz G, Rosset D, Barnea T J, et al.:Arbitrarily small amount of measurement independence is sufficient to manifest quantum nonlocality[J]. Physical review letters, 2014, 113(19): 190402.
	\bibitem{QRNG10}
	Yuan X, Cao Z, Ma X.:Randomness requirement on the Clauser-Horne-Shimony-Holt Bell test in the multiple-run scenario[J]. Physical Review A, 2015, 91(3): 032111.
	\bibitem{QRNG11}
	Yuan X, Zhao Q, Ma X.:Clauser-Horne Bell test with imperfect random inputs[J]. Physical Review A, 2015, 92(2): 022107.
	\bibitem{QRNG12}
	Li D D, Zhou Y Q, Gao F, et al.:Effects of measurement dependence on generalized Clauser-Horne-Shimony-Holt Bell test in the single-run and multiple-run scenarios[J]. Physical Review A, 2016, 94(1): 012104.
	\bibitem{QRNG13}
	Colbeck R, Renner R.:Free randomness can be amplified[J]. Nature Physics, 2012, 8(6): 450-453.
	\bibitem{QRNG14}
	Gallego R, Masanes L, De La Torre G, et al.:Full randomness from arbitrarily deterministic events[J]. Nature communications, 2013, 4(1): 1-7.
	\bibitem{QRNG15}
	Brandão F G S L, Ramanathan R, Grudka A, et al.:Realistic noise-tolerant randomness amplification using finite number of devices[J]. Nature communications, 2016, 7(1): 1-6.
	\bibitem{QRNG16}
	Ramanathan R, Brandão F G S L, Horodecki K, et al.:Randomness amplification under minimal fundamental assumptions on the devices[J]. Physical review letters, 2016, 117(23): 230501.
	\bibitem{CHSH}
	Clauser J F, Horne M A, Shimony A, et al.:Proposed experiment to test local hidden-variable theories[J]. Physical review letters, 1969, 23(15): 880.
	\bibitem{tilted}
	Huang X H, Li D D, Zhang P.:Effects of measurement dependence on tilted CHSH Bell tests[J]. Quantum Information Processing, 2018, 17(11): 291.
	\bibitem{Chain}
	Braunstein S L, Caves C M.: Information-Theoretic Bell Inequalities[J]. Physical Review Letters, 1988, 61(6):662.
	\bibitem{xin Bell}
	Collins D, Gisin N.:A relevant two qubit Bell inequality inequivalent to the CHSH Bell inequality[J]. Journal of Physics A: Mathematical and General, 2004, 37(5): 1775.
	\bibitem{9}
	Fine A.:Hidden variables, joint probability, and the Bell inequalities[J]. Physical Review Letters, 1982, 48(5): 291.
	\bibitem{family}
	Kaniewski J.:Weak form of self-testing[J]. Physical Review Research, 2020, 2(3): 033420.
	\bibitem{family2}
	Froissart M.:Constructive generalization of Bell’s inequalities[J]. Il Nuovo Cimento B (1971-1996), 1981, 64(2): 241-251.
	\bibitem{local}
	Rabelo R L D S.:On Quantum Nonlocality and the Device-independent Paradigm[D]. PhD thesis, 2013. 
	
\end{thebibliography}
\end{document}